\documentclass[reprint,amsmath,amssymb,aps,prl,showkeys,onecolumn]{revtex4-1}

\usepackage{graphicx}% Include figure files
\usepackage{float}
\usepackage{bm}% bold math
\usepackage{comment}
\usepackage[normalem]{ulem}
\usepackage{xr-hyper}
\usepackage[hidelinks]{hyperref}
\usepackage{color}
\makeatletter

\begin{document}
\preprint{APS/123-QED}

\title{Thin film Rupture from the Atomic Scale}

\author{Muhammad Rizwanur Rahman$^{1}$}
\email{m.rahman20@imperial.ac.uk}
\author{Li Shen$^{1}$}
\author{James P. Ewen$^{1}$}
\author{Benjamin Collard$^{1,2}$}
\author{D. M. Heyes$^{1}$}
\author{Daniele Dini$^{1}$}
\author{E. R. Smith$^{3}$}
\affiliation{
$^{1}$Department of Mechanical Engineering, Imperial College London, South Kensington Campus, London SW7 2AZ, United Kingdom, 
$^{2}$Department of Materials Science, Imperial College London, South Kensington Campus, London SW7 2AZ, United Kingdom,
$^{3}$Department of Mechanical and Aerospace Engineering, Brunel University London, Uxbridge UB8 3PH, United Kingdom
}

\begin{abstract}
The retraction of thin films, as described by the Taylor-Culick (TC) theory, is subject to widespread debate, particularly for films at the nanoscale. We use non-equilibrium molecular dynamics simulations to explore the validity of the assumptions used in continuum models, by tracking the evolution of holes in a film. By deriving a new mathematical form for the surface shape and considering a locally varying surface tension at the front of the retracting film, we reconcile the original theory with our simulation data to recover a corrected TC speed valid at the nanoscale.
\end{abstract}

\maketitle
{\textbf{{\textit{Introduction}}:}}
Thin film rupture is of premier interest in numerous physical and chemical transport processes~\citep{thiele2001dewetting,masson2002hole,de2004capillarity,deGennes2008,bird2010daughter,bico2015cracks, petit2015holes, poulain2018ageing}, ranging from disease transmission \citep{bourouiba2021fluid} to planetary scale environmental and oceanic science \citep{wu1981evidence,veron2015ocean}.
As such, the film rupture process has been investigated for over a century~\citep{marangoni1872monografia,ranz1959,chatzigiannakis2020breakup};
and remains an active area of debate.
Taylor~\citep{taylor1959dynamics} and Culick~\citep{culick1960comments} independently derived an equation for film retraction by considering that the growth of a nucleated hole of radius, $R$, is driven purely by the balance between the surface force, $F_\gamma$, and the inertia of the liquid mass, $m$, collected in a rim surrounding the hole, i.e., $F_{\gamma} = \frac{d}{dt}( m \dot{R} )$. Applying the initial condition at (time) $t=0$, $R=0$ for finite  $\dot{R}$, one obtains, upon integration, the Taylor-Culick (TC) speed,
\begin{equation}\label{eq:taylorspeed}
U_\mathrm{TC} =  \sqrt{{\phi\gamma}/{\rho h_0}}  
\end{equation} where $h_0$ is the thickness of the unperturbed part of the film with density $\rho$ and surface tension $\gamma$. In Taylor~\citep{taylor1959dynamics} and Culick~\citep{culick1960comments}, $\phi=2$, but later work use smaller values for thinner films \citep{mcentee1969bursting}. 
This surprisingly simple, yet elegant equation captures the overall retraction dynamics, and corrects Dupre's inappropriate energy balance assumption~\citep{dupre1869theorie}.
However, its accuracy is debatable, and it markedly fails for films thinner than $100\, \mbox{nm}$~\citep{mcentee1969bursting,evers1996,evers1997,planchette2019breakup,pierson2020revisiting} - this region constitutes the key focus of this letter. 

Taylor's resolution is based on a few key assumptions, (i) when rupture propagates, $h_0$ remains uniform except at the rim, and (ii) all liquid-mass of the expanding hole is collected inside the rim, i.e., $m =\rho \pi R^2 h_0 $. In a more formal derivation, Culick made the assumptions of uniform thickness and constant surface tension everywhere.
Whilst the assumptions leading to Eq.\,(\ref{eq:taylorspeed}) are not incorrect 
at the global scale, their local deviations become increasingly important as thickness decreases down to nanometers. 

Significant deviations from $U_\mathrm{TC}$ have been documented, especially in the limit of very thin films \citep{mcentee1969bursting,planchette2019breakup}, with a range of continuum based models attempting to explain these differences by consideration of the non-uniformity of film thickness \citep{keller1983breaking}, surface elasticity \citep{petit2015holes}, surface tension gradients \citep{de2022retraction} and viscous effects \citep{pierson2020revisiting, savva2009viscous}. 
With continuum models, any additional physics must be included by adding another conservation or balance law.
In this letter, molecular dynamics (MD) simulation is applied to fully understand the retraction process from the atomic scale. Any rupture event must start at the smallest scale, so MD is an ideal technique to fully understand the origins of this process.
It provides the full picture of a fluid down to the electrostatic atomic interactions; and the interface dynamics, surface tension and viscosity emerge as outputs from MD simulations. 
Hence, many fluid properties have been successfully studied by MD, a previous letter in this journal, \citet{koplik2000molecular} used MD to show the origins of dewetting, alongside the studies of film breakup \citep{koplik1993,Zhao_et_al19}, boiling and nucleation \citep{diemand2014direct} and a range of cases 
\citep{kadau2010atomistic,todd2017nonequilibrium}.
\citeauthor{kono2014elevation} \citep{kono2014elevation} studied a two-dimensional film breakup with MD, and observed temperature rise due to rapid rupture which would cause a decrease in surface tension.
\begin{figure*}[htbp!]
\includegraphics[width=0.98\linewidth]{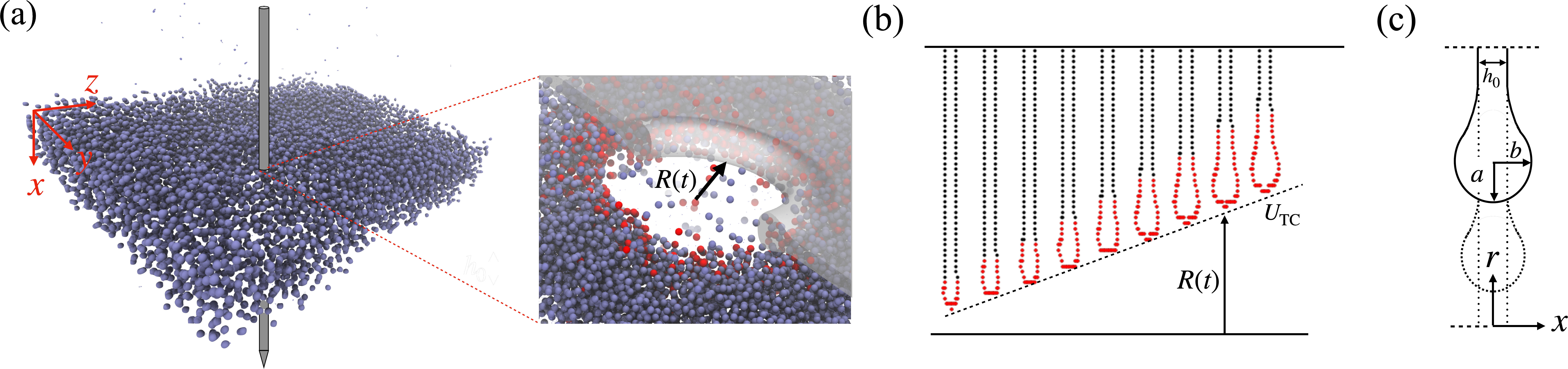}
\caption{\label{fig:schematic}(a) Computational domain considered for this study. A hole is nucleated at the center of the film in the $yz$ plane. A cylindrical needle schematically shows the hole-poking process.
Zoomed view of the hole: atoms in red denote the atoms initially in the hole at $t=0$ that are displaced; faint schematic shows an idealization of the rim surrounding the hole which borders the unperturbed film.
(b) Radial average of a retracting film over time: $R(t)$ is measured from the center of the hole to the tip of the rim. (c) Simplified schematic of panel (b), film thickness, $h_0$, is measured at the unperturbed part of the film.}
\end{figure*}

In this work, we model the full three-dimensional film breakdown and systematically outline 
all possible contributions from the atomic scale. Using novel local pressure measurement techniques, interface tracking and dynamic time-evolving mappings, local surface tension is calculated. This uncovers the full picture of surface force and interface shape, putting the thermal effects in context and exposing other key factors contributing to the dynamics. 
The TC model does not take into account any surface-species and their contributions to the surface properties of the film
~\citep{tammaro2021flowering,manikantan2020surfactant}. By construction, the film investigated in this work is pure in nature, and hence, closest to the TC model.
Present work thereby tests this model by removing external factors and shows that only when certain atomic-scale corrections are made, TC equation can successfully predict the retraction process.

\textit{\textbf{Simulation setup:}} Films investigated in this study were modeled using a 
Lennard-Jones (LJ) fluid simulated in the extensively validated and verified Flowmol MD code~\citep{smith2013coupling}. 
The initial simulation domain was a cubic box of dimensions $L_x = 76.19$ and $L_y = L_z =609.56$ in LJ units.
These dimensions correspond to $\sim 207$\,nm wide films with $h_0<5$\,nm.
Rest of this letter reports all quantities in LJ units.
The middle $20\%$ of the simulation box in the $x$ direction was designated as liquid and setup with a density, $\rho\approx0.7$ and  the remaining as vapor with a lower density, $\rho=0.01$.
Schematic in Fig.~\ref{fig:schematic}\,(a) shows the system modelled in this study where all atoms ($\sim3.54$ million) are identical, forming a central liquid film coexisting with surrounding vapor at equilibrium.
After equilibrating the film at $T=0.78\pm0.03$, a hole of initial radius, $R_0$ was induced in the otherwise stable film, and $R$ was measured over time. 
The inset in Fig.~\ref{fig:schematic}\,(a) (also {\color{black}SI Fig.\,1 \citep{mrr2022SI}}) shows the liquid atoms surrounding an expanding hole.
Displaced atoms (in red) from the hole collect around the hole to form a rim.
The radial-averaged film extracted from MD data, in panel (b), shows its temporal evolution - with simplified schematic in panel (c). 

The hole creation technique may invoke slight differences to the initial conditions, but any difference rapidly ceases~\citep{frankel1969bursting}.
We confirm this by applying two separate techniques to create holes, by 
(i) deleting atoms from a circular region of radius, $R_0$, or (ii) applying a small radially decaying force for a short time that models poking a film.
Henceforth, these two types will be referred to as cut- and poked-holes, respectively. 
The growth of spontaneously nucleated holes (by allowing holes to form naturally due to thermal fluctuations) were found to give similar dynamics, albeit spontaneous holes take considerably longer time to nucleate.
Each reported case in this letter was averaged over at least three separate ensembles. The short-time growth, being more susceptible to thermal noise, was averaged over ten separate ensembles, further details in {\color{black} S1 \citep{mrr2022SI}}. 

Surface deformations arising from thermal motion increases the likelihood of spontaneous rupture \citep{vrij1966,vrij1968,Ivanov1970}. Therefore,
to find a stable, and computationally tractable case, films with a range of thicknesses were tested without applying any external perturbation. A thickness, $h_0 = 14$ was selected as films thinner than this were observed to break spontaneously before reaching a constant retraction rate. Other computationally affordable thicknesses were also examined ({\color{black}S1.1 \citep{mrr2022SI}}).
\begin{figure*}[htbp!]
\includegraphics[width=0.95\linewidth]{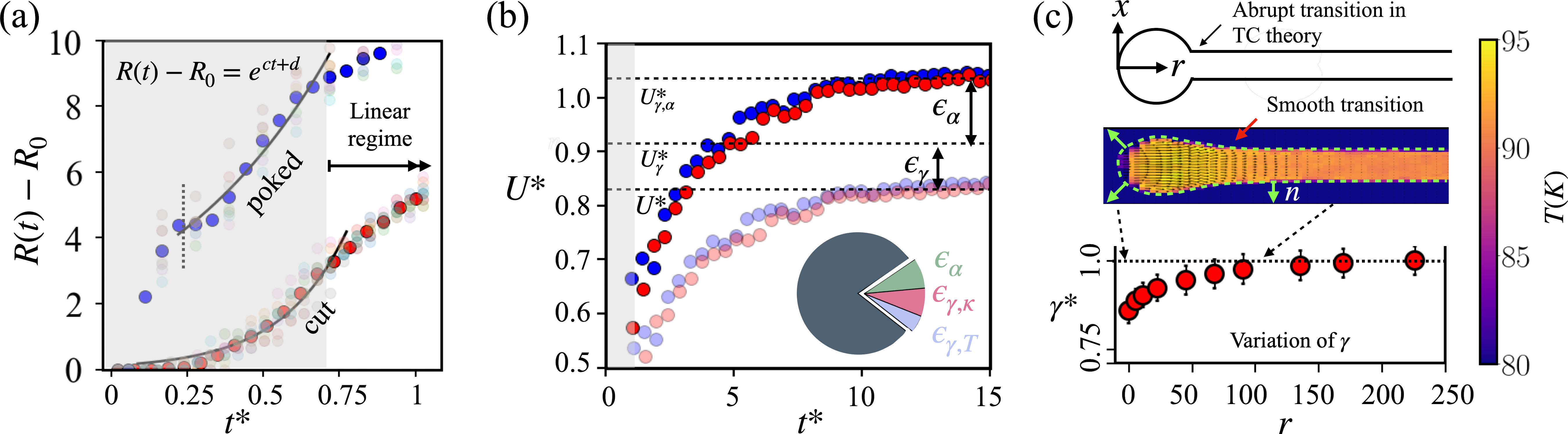}
\caption{\label{fig:Ulong}(a) Short-term nonlinear retraction of film for cut- and poked-holes with initial thickness is $h_0\sim 14$, time is non-dimensionalized by the inertial time scale, i.e., $t^* = t/\sqrt{\rho h_0^3/2\gamma}$. Faint and solid symbols denote, respectively, data from individual simulation, and data averaged over 10 simulations. Solid line is an exponential fit to the data of the form: $R(t) - R_0 = e^{\alpha t + \beta}$.
(b) Retraction velocity (blue: poked-, red: cut-holes) normalized by the TC speed, $U^*=U/U_\mathrm{TC}$ with $2\%$ standard deviation.
Faint symbols denote cases where the classical $U_\mathrm{TC}$ is used, whereas, 
solid symbols denote cases where $U_\mathrm{TC}$ is corrected considering the local surface tension ($U^*_\gamma$), and the existence of a transition regime ($U^*_{\gamma,\alpha}$). Pie-chart in the inset shows various corrections arising from the change of $\gamma$ due to temperature, $\epsilon_{\gamma,T}$ and curvature, $\epsilon_{\gamma,\kappa}$, and due to the transition regime, $\epsilon_\alpha$. In panels a and b, the non-linear growth region is shaded.
(c) Schematic in the top-panel shows an idealized retracting film. 
The mid-panel sheds molecular insights where a transition region is present between the rim and the unperturbed film. Green arrows show the directions normal ($n$) to the local film surface. Color map shows radial-averaged instantaneous temperature field with velocity vectors (black arrows) overlaid.
Bottom panel shows the local to global (time-averaged) surface tension ratio, ($\gamma^*=\gamma_r/\gamma$) against distance, $r$ from the tip of the retracting film.}
\end{figure*}

{\textbf{{\textit{Results and discussions}}}}:
We chart the hole-growth as predicted by MD and compare to experiments, CFD and TC theory, starting from initial hole formation, through a short-time exponential regime before eventually reaching a constant speed which can be shown to match TC when we account for local changes in shape and in surface tension.

{\textit{Initial hole formation}}: For systems at macroscopic length scale, growth of holes with an initial radius, $R_0<h_0$ is not energetically favoured, i.e., an open hole closes~\citep{taylor1973making}.
Hence, earlier experiments or numerical investigations~\citep{mcentee1969bursting,pandit1990,savva2009viscous, sanjay_sen_kant_lohse_2022} on sheet retraction considered an initial hole dimension satisfying, $R_0 \gg h_0$. 
Additionally, continuum assumptions require long-wave approximations to describe the rupture process which is valid for $h_0$ much larger than the atomic scale~\citep{vaynblat2001rupture}.
However, in the limit of a few nanometers thickness (essentially Newtonian black films), the disjoining force acts as a destabilizer opposing the stabilizing effect of surface tension, and therefore, becomes responsible for the puncture of
the film~\citep{neel2018,villermaux2020fragmentation}. 
At this scale; MD is the only appropriate tool to scrutinize the hole opening process.
It was observed that $R_0\gg h_0$ no longer remains a requirement for holes to expand ({\color{black} S1.2 \citep{mrr2022SI}}).
This agrees with the seminal study of~\citet{koplik2000molecular}, where the authors reported that for any dry patch larger than a few  molecular diameters, the separated molecules suffer from mutual interactions, and the hole always grows.

{\textit{Short-term retraction}}:
At the onset of retraction, an exponential growth for a short period of time is observed, which converges to a linear terminal speed at later times.
Fig.~\ref{fig:Ulong}\,(a) portrays this short-term hole expansion for cut- and poked-holes. In agreement with the observations in \citep{debregeas1995viscous,masson2002hole}, it is evident that the early-time hole growth shows an exponential behavior, i.e., $R(t) - R_0 = \mathrm{exp}\,(c t + d)$, where $c$ and $d$ are fitting parameters. 
While \citeauthor{debregeas1995viscous}\,\citep{debregeas1995viscous} ascribed this short-term growth to the viscoelastic effects, it was confirmed to be a generic feature of circular hole retraction\,\citep{savva2009viscous}, and was also accredited to the initial shape of the puncture~\citep{roth2005evidence,savva2009viscous}.

{\textit{Taylor-Culick retraction}}:
Beyond the short-term exponential regime, the retraction velocity approaches a plateau - the constant TC speed \citep{savva2009viscous}. This is shown by the faint symbols in Fig.\,\ref{fig:Ulong}\,(b) for
cut- (red) and poked-holes (blue). 
Retraction speed, $U$ is measured over $50$ consecutive time-steps, and is non-dimensionalized by $U_\mathrm{TC}$ from Eq.\,(\ref{eq:taylorspeed}) with $\phi=2$, i.e., $U^* = U/U_\mathrm{TC}$. 
Evident from Fig.\,\ref{fig:Ulong}\,(b), $U_\mathrm{TC}$ is appreciably faster than the measured speed ($U^*\sim0.8$), that is MD predicts a speed which is only 80\% of the Taylor-Culick prediction.
This is not unexpected, and documented in the literature~\citep{mcentee1969bursting,muller2007experimental,trittel2013rim,petit2015holes,tammaro2021flowering}.
Notably,~\citet{mcentee1969bursting} studied the retraction of soap-films with a wide range of thicknesses.
In their experiments, the retraction speeds for films of thickness smaller than $100\,\mbox{nm}$ were found to be substantially slower than $U_\mathrm{TC}$, requiring $\phi\sim0.7$.
$U_\mathrm{TC}$ is rather closer to our measured speed 
%($20\%$ higher than present study vs. $60\%$ higher than \citeauthor{mcentee1969bursting}'s experiments) and 
requiring $\phi\sim 1.35\pm0.05$ for $h_0\sim4.75$\,nm. 
However, having the effects of surfactants and visco-elasticity~\citep{vrij1968} completely eliminated in the present study (by considering pure film), why the TC theory still fails to estimate the retraction rate is explored now.

A number of factors, i.e., temperature variation due to internal heat generation, curvature effects and viscous contribution~\citep{pierson2020revisiting} might play a critical role in slowing down the retraction. These are often characterized by the Ohnesorge number, $\mathrm{Oh}\,=\mu / \sqrt{\rho \sigma h_0}$, which is $\sim 0.8$ for the present case.

{\textit{Surface tension - temperature effects}}:
Adjusting the value of $\phi$ marries the theoretical estimate with the measured speed, but it does not necessarily explain the underlying physics responsible for the slower retraction.
This disagreement was attributed to the disturbance caused by the presence of a precursor ahead of the rim arising from surface tension gradient due to surface shrinkage and surfactant concentration~\citep{mcentee1969bursting, mysels1971}.
~\citet{frankel1969bursting} inferred that the slow growth-rate should be due to a large variation of the local surface tension at the film-front.
Although such local fluctuation is not unusual~\citep{moody2001curvature, wen2021molecular}, capturing these with continuum models is difficult, and the only MD study is the quasi-2D case of \citet{kono2014elevation} who reported  considerable temperature rise ($>5K$) in the liquid surrounding the hole for $h_0\sim5$\,\mbox{nm}.

In contrast to 2D systems, fully 3D simulations are computationally more expensive, but avoid the well-known deficiencies of 2D fluids, such as divergent transport coefficients~\cite{irving1950}.
One expects the behavior of an interface to be dominated by the number of interacting neighboring molecules - which is much smaller in 2D. Therefore, a 2D interface would be overly susceptible to rupture~\cite{koplik1993}.
Even with periodic boundary conditions, the strong system-size dependency of the diffusion coefficient cannot also be ignored \citep{yeh2004system}.
Present study therefore minimizes these effects by simulating a fully three-dimensional (3D) system which is $\sim8$ times larger (in terms of particle number) than the 2D system of \citep{kono2014elevation}.
A warmer region surrounding the expanding hole is observed, and the temperature increase is rather less pronounced, $2.9$K vs. $>5$K, {\color{black} see S2.2 \citep{mrr2022SI}}. Such an increase in temperature constitutes around $5-7\%$ drop in $\gamma$.
This is, of course, non-negligible; but is not sufficient to explain the slower retraction. 

{\textit{Surface tension - curvature effects}}:
The retraction process is mainly driven by the tangential component of the pressure tensor, while the normal component is responsible for the radial growth of the rim.
Resolving the pressure tensor spatially along the film profile in the local tangential and normal directions ({\color{black} S2.2 \citep{mrr2022SI}}) reveals a noticeable depression of the tangential pressure.
From these constituent components, 
$\gamma$ can be locally calculated using the Kirkwood-Buff definition~\citep{kirkwood1949statistical}, i.e., $\gamma_r = \frac{1}{2}  \int_{-\infty}^{\infty} \left[P_{xx} (r,x) - P_{rr} (r,x)\right]dx$, where $r$ denotes the radial distance from the film tip, subscripts $xx$ and $rr$ denote direct pressure along the x and r axis, respectively. 
A full rotation and mapping of the pressure tensor using a radially-varying normal vector does not give a substantially different surface tension (see {\color{black} S2.2 \citep{mrr2022SI}}).
The tension acts as a `pulling force' in Taylor-Culick formulation, purely in the direction of retraction, seen from the Kirkwood-Buff with $P_{rr}$ the main contribution to $\gamma$. 
The bottom-panel in Fig.\,\ref{fig:Ulong}\,(c) shows the ratio of the local to global surface tension with increasing $r$.
Remarkably, this gives evidence of a substantial drop in $\gamma_r$ ($\sim15\%$) at the tip of the retracting film ($r=0$), and $\gamma_r \to \gamma$ far from the rim.
While the temperature rise explains a $\sim 5-7\%$ drop ($\epsilon_{\gamma,T}$) in $\gamma$, the remaining can be attributed to surface curvature and shrinkage, $\epsilon_{\gamma,\kappa}$, so $\epsilon_{\gamma}=\epsilon_{\gamma,T}+\epsilon_{\gamma,\kappa}$.
Accounting for this local variation, an improvement $\epsilon_\gamma = 1 - \sqrt{\gamma_{r = 0}/\gamma}$ is obtained; and for the present case, using the surface tension averaged over the film front
in Eq.\,(\ref{eq:taylorspeed}) results in agreement with the measured speed within only a $\sim8\%$ deviation, shown by the dashed line denoted by $U^*_\gamma$ in Fig.~\ref{fig:Ulong}\,(b).
These corrections, however, will diminish for thicker films as the variations of temperature \citep{kono2014elevation} and curvature become less pronounced ({\color{black} S2.2 \citep{mrr2022SI}}).

{\textit{Transition regime}}: The top panel in Fig.\,\ref{fig:Ulong}\,(c) shows an idealized picture of film retraction upon which the TC formalism is based on. This simplified description requires the liquid from the hole to be collected in a spherical rim bordering on the unperturbed film.
Nature will rectify such abrupt discontinuity through the effects of viscosity~\citep{rayleigh1880stability}.
Contrasting Taylor's assumption of a sudden transformation, present study rather agrees with experimentally~\citep{lhuissier2009soap,petit2015holes} and numerically~\citep{koplik2000molecular} observed transition, shown in the mid-panel of Fig.\,\ref{fig:Ulong}\,(c). 
Previous studies reported an `aureole-shaped' regime for films of liquid with surfactants, and attributed this to surface tension gradient, and surface elasticity ~\citep{de2022retraction,lhuissier2009destabilization,petit2015holes}. Existence of a similar `extended rim' in the present study - where the film is made of pure liquid - indicates a gradient of the surface force, but caused by curvature effects (geometry) and viscosity\,\citep{savva2009viscous}.

The classical formulation discounts any possibility of a transition regime between the rim and the undisturbed film. Our results confirm that the rim is followed by a transition regime (Fig \ref{fig:schematic}\,b).
This can be explored through a mass balance argument by assuming an elliptic blob, of aspect ratio $\alpha=b/a$, connected by a smooth transition region, of length $L$, to the neck of width $h_0$.
The length, $L$ is given by,
\begin{equation}\label{eq:aureole_length}
    L={\pi h_0} (\beta^2-1)/ {2\alpha}
\end{equation}
where, $\beta = 2b / h_0$ ({\color{black}see S2.3\,\citep{mrr2022SI}} for full derivation). 
Transition must take place over the finite length, $L$~\citep{frankel1969bursting} - which we confirm by comparing $L$ to our simulations ({\color{black}SI Fig.\,6(b) \citep{mrr2022SI}}), thus disproving, for thinner films, the assumption of an unperturbed film bordering on the rim.
As Eq.\,(\ref{eq:aureole_length}) suggests, the transition length becomes more prominent with thinner films, and disappears when $b\sim h_0/2$ corresponding to viscous film retraction \citep{brenner1999bursting,savva2009viscous}. 

The MD data indicates that the normal pressure difference of the surface is minimal at
the point where the rim meets the transition region. In order to satisfy the Young-Laplace equation, it follows that the absolute value of the curvature must also be minimized.
Under these assumptions, the atomistic form of Eq.\,(\ref{eq:taylorspeed}) can be expressed as 
\begin{equation}\label{eq:transition}
   U= \left(1+\alpha^2\right)^{-1/4} U_\mathrm{TC}.
\end{equation}
For simplicity, assuming a spherical rim, i.e., $\alpha=1$ yields a surface-shape correction factor, $\epsilon_\alpha = 2^{-1/4}\approx 0.84$. 
When $\epsilon_\alpha$ is accounted for, in addition to $\epsilon_\gamma$, the corrected TC model ($U^*_{\gamma,\alpha}$) shows close agreement with the MD results.
This is shown by the solid symbols in Fig.~\ref{fig:Ulong}\,(b), and as evidenced by the figure, $U^*_{\gamma,\alpha} \sim 1$. 

There remain a few other aspects which plausibly affect the dynamics of film retraction. Any mutual dependency of the two correction factors developed in this study cannot be ruled out.
The growth of the blob is non-negligible as compared to $\dot{R}$. In situations like this, \citet{pierson2020revisiting} and others  \citep{savva2009viscous,de2022retraction} argued that the velocity within the blob is non-uniform; the velocity vectors overlaid on the temperature field in Fig.\,\ref{fig:Ulong}\,(c) confirm this hypothesis.
The effect of this velocity would contribute to the departure of the blob shape from being circular and therefore, is assumed to be included in the surface shape correction, $\epsilon_{\alpha}$. The effect of evaporation, however, in agreement with previous study~\citep{kono2014elevation}, was found to be negligible.\\

{\textbf{{\textit{Conclusion}}:}}
This study examines the Taylor-Culick theory for film retraction at atomic scale using large-scale MD simulations.
Similar growth rate of holes made by two separate methods, either poking or cutting, confirms that the dynamics is independent of hole creation mechanism.
This then allows a systematic study, where we repeatably show a short term exponential growth period followed by a constant velocity regime over an ensemble of cases - both behaviours observed in the experimental literature.
A Lennard-Jones model is used to remove viscoelastic effects so the various driving contributions to the growth dynamics can be exposed.
The theoretical conjecture of a constant terminal retraction rate is accurate at this scale, however, we show that estimating the `true' retraction speed requires corrections for (i) the local variation of surface tension arising from surface curvature and  temperature rise, and, (ii) the existence of a transition regime between the rim and the unperturbed film - both of these factors were ignored in the classical formulation.
Surface tension, viscosity and interface shapes are outputs of MD simulations and we observe how they contribute to the dynamics, and quantify their impact.
While their contributions are marginal for macroscopic films, these are the key reasons for the failure of the classical theory as film thickness is reduced below $100$\,nm.

This letter provides compelling evidence that the pioneering theory of Taylor and Culick, which in its original form fails at the atomic scale, can be corrected to accurately predict the film retraction speed.

\begin{acknowledgments}
M.R.\,R. thanks Shell and the Beit Trust for PhD funding through a Beit Fellowship for Scientific Research. L.S. thanks the Engineering and Physical Sciences Research Council (EPSRC) for a Postdoctoral Fellowship (EP/V005073/1). J.P.E. was supported by the Royal Academy of Engineering (RAEng) through their Research Fellowships scheme. B.C. was supported by Shell and the EPSRC via an iCASE PhD studentship (EP/T517690/1). D.D. acknowledges a Shell/RAEng Research Chair in Complex Engineering Interfaces and the EPSRC for an Established Career Fellowship (EP/N025954/1).
\end{acknowledgments}

\end{document}

\preprint{APS/123-QED}

\title{Supplementary Material \\ Thin Film Rupture from the Atomic Scale}

\author{Muhammad Rizwanur Rahman$^{1}$}
\email{m.rahman20@imperial.ac.uk}
\author{Li Shen$^{1}$}
\author{James P. Ewen$^{1}$}
\author{Benjamin Collard$^{1,2}$}
\author{D. M. Heyes$^{1}$}
\author{Daniele Dini$^{1}$}
\author{E. R. Smith$^{3}$}
\affiliation{
$^{1}$Department of Mechanical Engineering, Imperial College London, South Kensington Campus, London SW7 2AZ, United Kingdom, 
$^{2}$Department of Materials Science, Imperial College London, South Kensington Campus, London SW7 2AZ, United Kingdom,
$^{3}$Department of Mechanical and Aerospace Engineering, Brunel University London, Uxbridge UB8 3PH, United Kingdom
}

\maketitle
\tableofcontents
\newpage 
\begin{figure*}[tbp]
    \centering
    \includegraphics[width=0.85\linewidth]{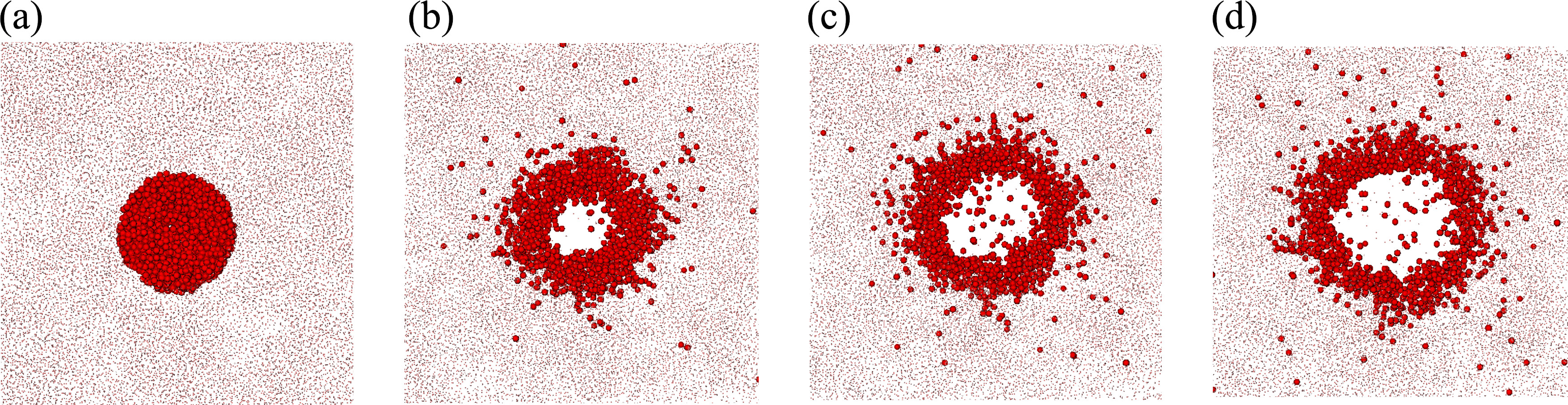}    
    \caption{\label{app:fig:0}Growth of a spontaneously nucleated hole, (a) before nucleation, (b-d) as the hole expands. Atoms displaced from within the hole of panel (d) are colored in red in each of the panels, remaining liquid atoms are faintly drawn. The displaced atoms are collected in the rim surrounding the hole as it grows.}
    \label{app:fig:hole-growth}
\end{figure*}

\section{1.\,Methodology}\label{app:sec:methodology}
The middle $20\%$ of the simulation box was initialized with LJ particles in the liquid phase and the remaining of the simulation domain was designated as vapor. 
To ensure that the periodic boundary does not affect the statistics, we have tested the growth dynamics for film with lateral dimensions (in LJ units) of (i) $304.78\times304.78$ with $\sim0.88$ million particles, (ii) $609.56\times609.56$ with $\sim3.54$ million particles, and (iii) $ 1219.12\times1219.12$ with $\sim14.1$ million particles.
For (i), the hole soon reaches the film boundary giving a shorter time-window for the rupture speed to reach a plateau, and (iii) demands considerably higher computation time while producing similar statistics as of (ii). As such, all reported cases are simulated for the lateral dimension of (ii).  
A shifted LJ potential, as in Eq.\,(\ref{eq:ljst}) is used to model the system, expressed mathematically, 
\begin{align}
        U_{ij} (r) = \left\{
            \begin{array}{ll}
                U_{ij} & \quad r \leq r_c \\
                0 & \quad \mathrm{otherwise}.
            \end{array}
        \right.
        \label{eq:ljst} 
\end{align}
where, $r_c$ is the cut-off distance and,
\begin{displaymath} 
    U_{ij} (r) = 4 \epsilon_{ij} \left [ \left( \frac{\sigma_{ij}}{r_{ij}} \right)^{12} -\left( \frac{\sigma_{ij}}{r_{ij}} \right)^6 \right]
\end{displaymath}
here, $U_{ij}$ is the potential between particle $i$ and $j$ that are located at a distance $r_{ij}$, $\epsilon_{ij}$ is the dispersion energy, and $\sigma_{ij}$ is the characteristic length scale.
Periodic boundary conditions are applied in all three Cartesian directions.

The system was allowed to equilibrate until a stable liquid$-$vapor coexistence was established.
This was run in the canonical (NVT) ensemble using a No\'{s}e-Hoover thermostat
\citep{nose1984molecular,hoover1985canonical}. 
Once equilibrated, a hole is created at the center of the $YZ$ plane. Two separate hole-making techniques were employed, (i) cut-holes: where molecules within a circular region of radius $R_0$ were randomly deleted to achieve a target vapor density, $\rho_{h,0}$ within the hole, and (ii) poked-holes: where an external force, as in Eq.\,(\ref{eq:force_ext}), was applied to mimic the popular experimental procedure of puncturing a film~\citep{savva2009viscous,bird2010daughter,oratis2020new}. 

\begin{align}
        F_\mathrm{ext.} = \left\{
            \begin{array}{ll}
                 F \left( 1 + \mathrm{e}^{-1/r^2} \right) & \quad r \leq R_0 \\
                0 & \quad \mathrm{otherwise},
            \end{array}
        \right.
        \label{eq:force_ext} 
\end{align}
where, $R_0$ is the target initial radius of the hole. In case of poked holes, the forcing time was only a few hundred steps to confirm that a stable hole is punctured, the system was then run in the micro-canonical (NVE) ensemble.
For spontaneous holes, the films were run in NVE ensemble until a hole naturally nucleates due to atomic-scale thermal fluctuations and expands enough to reach constant terminal velocity. Fig.\,\ref{app:fig:0} (a-d) show the growth of spontaneously nucleated hole over time.\\

\begin{figure*}[tbh]
\includegraphics[width=0.65\linewidth]{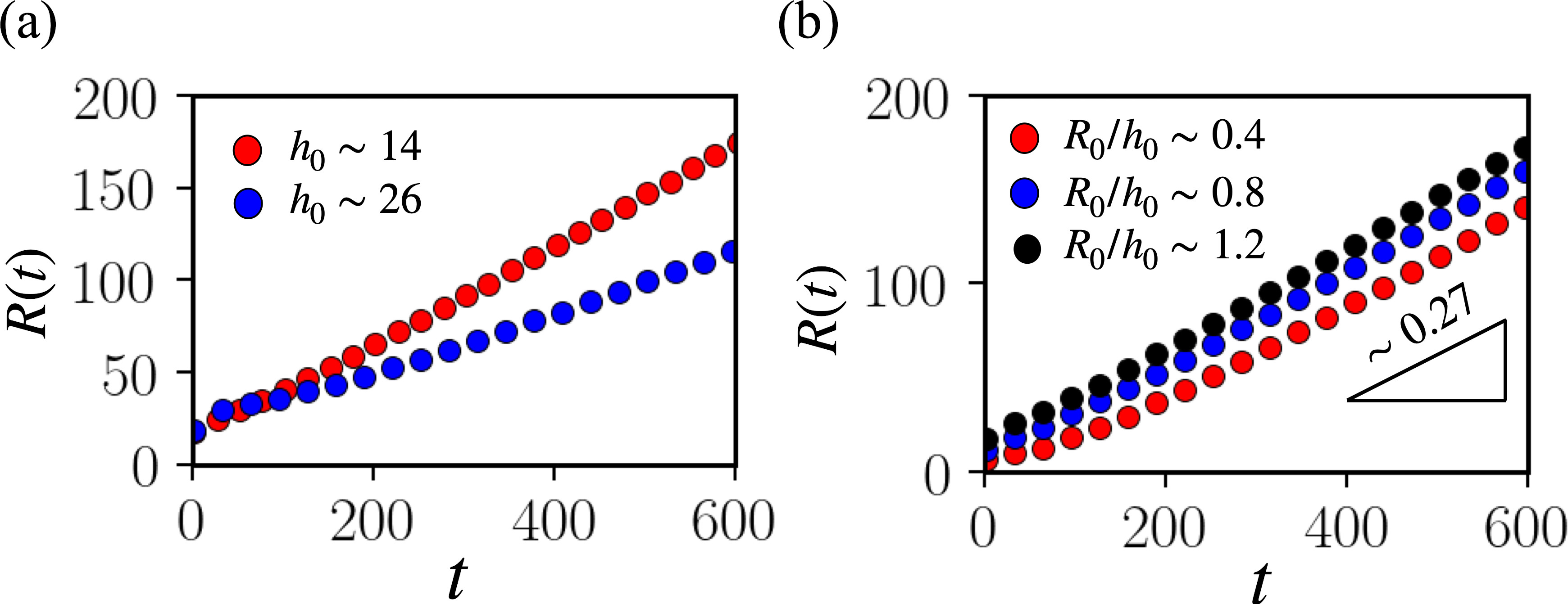}
\caption{\label{app:fig:1}(a) The growth rate (Taylor-Culick speed) of an expanding hole decreases for thicker films. (b) Terminal retraction rate is independent of the initial hole radius.}
\end{figure*}

In order to reduce the computational cost in simulating the relatively large systems, a cut-off radius, $r_c=2.5$ is used. Although a smaller cut-off radius results in a smaller surface tension; accordingly the retraction rate should similarly scale down. However, this should not impact the overall retraction dynamics. 
Near-zero densities are observed inside the hole  which attain the liquid density after a slight jump at the shell (rim) surrounding the hole. While cutting the hole, a target density $\rho_{h,0}=0$ was specified within a circular region of radius, $r=R_0$. The effect of the target density $\rho_{h,0}$ is tested by setting it to different values in the range of $[0, 0.025]$, however no noticeable differences in the film density, or in the overall dynamics were observed. In all the cases reported, both cut-holes and poked-holes, the average density of the liquid film is found to be $0.7\pm0.01$. 

\subsection{1.1 Selection of Film Thickness}\label{app:sec:film-thickness}
The inherent instabilities in ultra-thin films, owing to the long range disjoining forces and the mechanical or thermal perturbations~\citep{debregeas1998life}, make the examination of the rupture process challenging. 
To capture the growth dynamics of the induced holes,
we first examined films with a wide range of thicknesses to ensure that the selected film does not rupture on its own.
We observed substantial increase of film stability (i.e., exponential increase of natural rupture time) with increasing thickness. To study the hole-growth dynamics, we choose an initial film thickness, $h_0 \sim 14$, and observed that such a thickness allows sufficient time when the film would not spontaneously break. As such the growth dynamics of the induced hole can be considered free from any effects otherwise caused by the instability of the film.

\subsection{1.2 Effect of Initial Conditions}\label{app:sec:initial-conditions}
With all other parameters unchanged, Taylor-Culick equation (Eq. 1 in the manuscript) suggests an inverse dependency of the retraction speed on the film thickness, $h_0$, i.e., $U \propto 1/\sqrt{h_0}$. This is evident in Fig.\,\ref{app:fig:1}\,(a), which plots $R(t)$ with time for $h_0\sim 14$ and $\sim 26$. 
The effect of the initial radius of the induced holes are examined with different radii, i.e., $R_0/h_0 \sim 0.4, 0.8$ and $1.2$ in Fig.\,\ref{app:fig:1}\,(b). Irrespective of $R_0$, holes attain same terminal speed. Indeed, for lower $\mathrm{Oh}=\mu/\sqrt{\rho \gamma h_0}$, the relatively shorter timescale involved to approach the terminal speed minimizes any effect of initial radius~\citep{savva2009viscous}.
 
\section{2. Corrections to Taylor-Culick Speed}\label{app:sec:TCderiv}
\subsection{2.1\,Classical Taylor-Culick Speed}
The lone acting force that drives the film retraction process, i.e., the surface force, $F_{\gamma}$ must be balanced by the inertia of the mass, $m$ collected in the rim. When this dynamic balance is attained, the film retracts with a constant velocity, $U_\mathrm{TC}$ (although, for smaller $\mathrm{Oh}$, this assumption of uniform velocity inside the blob becomes invalid~\citep{pierson2020revisiting}). The surface force, $F_\gamma$ can be expressed as:
\begin{equation}\label{eq:force}
    F_\gamma = 4\pi R(t) \gamma
\end{equation}
On the other hand, the change of momentum, $P$ of the accumulated mass is:
\begin{equation}\label{eq:momentum}
  \frac{dP}{dt}  = U_\mathrm{TC}\frac{dm_h}{dt} 
\end{equation}
As the liquid from the hole is accumulated in the rim,  the continuity equation gives: ${dm_h}/{dt} =  2\pi \rho R(t) h_0  U_\mathrm{TC}$. Using this expression and equating (\ref{eq:force}) and (\ref{eq:momentum}) yields:\,$4\pi R(t) \gamma =  2 \pi \rho R(t) h_0 U_\mathrm{TC}^2$,
and the Taylor-Culick speed can be expressed as:
\begin{equation}
   U_\mathrm{TC} = \sqrt{\frac{2\gamma}{\rho h_0}}.
\end{equation}

\subsection{2.2 Correction for Local Surface Tension}
\begin{figure}[th]
\centering
\includegraphics[width=0.70\linewidth]{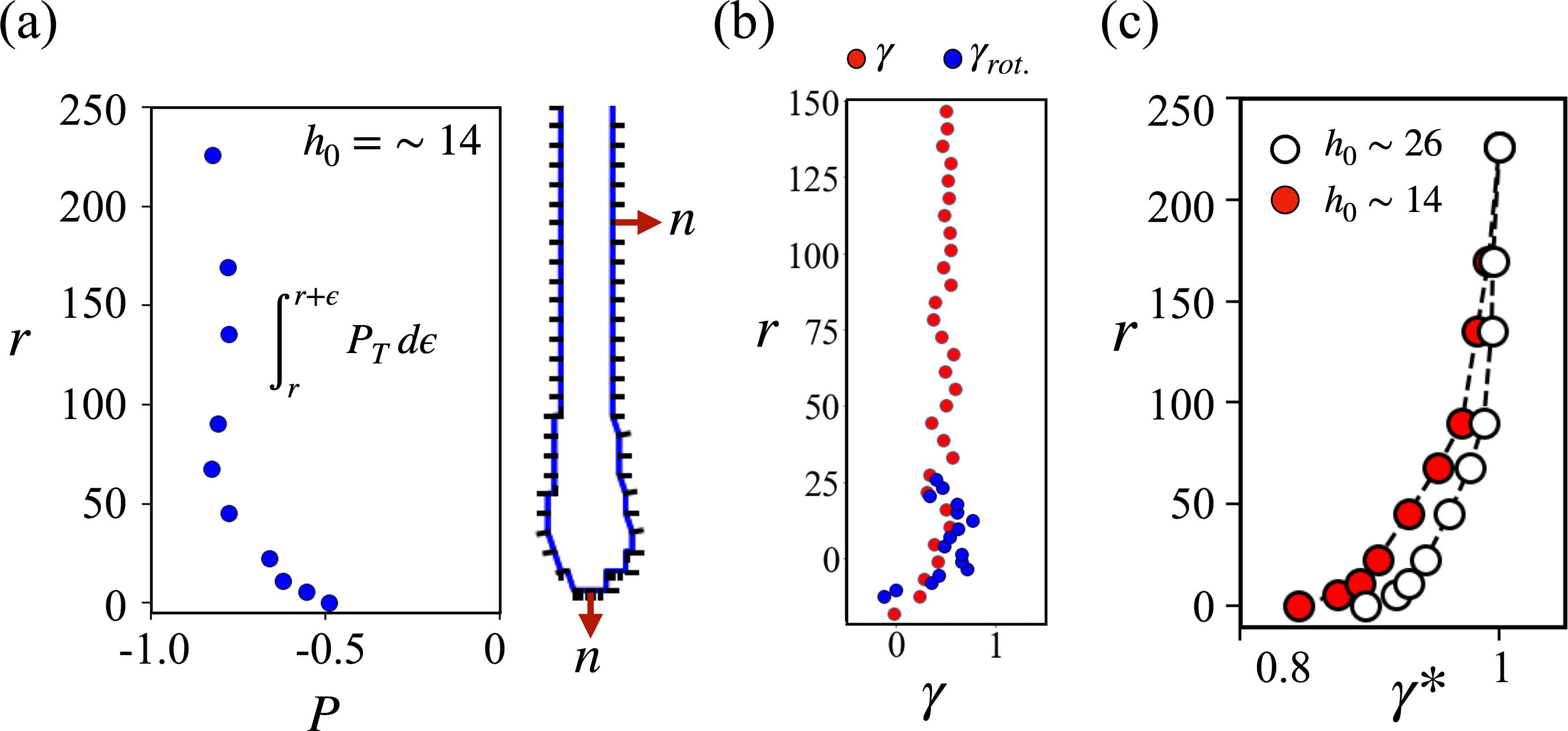}
\caption{\label{app:fig:local}(a) Tangential component of pressure along the radial direction of a film of thickness, $h_0 \sim 14$. Schematic at the right shows local normal directions to the surface. (b) Rotation of the pressure tensor along the local normal and tangential directions at the film front gives broadly similar surface tension (blue symbols) as of the un-rotated surface tension (red symbols). (c) The drop in local surface tension at the film front for a relatively thicker film ($h_0 \sim 26$) is less pronounced than that of a thinner film ($h_0 \sim 14$).}
\end{figure}
Following \citeauthor{kirkwood1949statistical}\,\citep{kirkwood1949statistical} definition, surface tension of the film is obtained from the Hulshof integral~\citep{hulshof1901direct} as:
\begin{displaymath}\label{Eq:surfTenMec}
    \gamma_r = \frac{1}{2}  \int_{-\infty}^{\infty} \left( P_{xx} (r,x) - P_{rr} (r,x) \right) dx.
\end{displaymath}
Here $P_{xx} (r,x)$ is the normal pressure at location $r$, where $x$ denotes the direction normal to the film surface at $r$; $P_{rr} (r,x)$ is the tangential pressure at $r$.
The time-averaged surface tension is found to be $0.51$ which is in excellent agreement with the literature~\citep{mecke1997molecular} for similar cut-off radius.
In order to determine any local variation to the surface force, we inspect two possible root causes, (i) surface curvature, and (ii) local temperature - these are elaborated below.

\noindent The Cartesian coordinate system is first transformed to radial coordinates. Later, the tangential components of the pressure tensor are transformed as:
\begin{align}
\begin{array}{ll}
    P_{t1} &= P_{xx}\,\mbox{cos}^2\theta 
          + P_{yy}\, \mbox{sin}^2\theta 
          + P_{xy}\,\mbox{sin} 2\theta \\
    P_{t2} &= P_{xx}\,\mbox{sin}^2\theta 
          + P_{yy}\, \mbox{cos}^2\theta  
          - 2 P_{xy}\,\mbox{sin} \theta \mbox{cos}\,\theta
\end{array}          
\end{align} 

The normal forces acting around the rim should cancel each other. To identify any local variation, a strip of width $\epsilon$ 
is defined over which the tangential and normal pressure are integrated. The time averaged (local) tangential pressures obtained in this fashion are plotted in Fig.\,\ref{app:fig:local}\,(a) which shows the noticeable depression of the tangential pressure at the tip as compared to the far-field. 

{\textbf{Rotation of pressure tensor:}} A rotation of the pressure tensor to get true normal and tangential pressure at each point on the surface,  together with an integral along the local surface normal $\textbf{n}(r)$ at each point on the interface, would give $\gamma_n = \frac{1}{2}  \int_{-\infty}^{\infty} \left[\boldsymbol{P} \cdot \textbf{n} - \boldsymbol{P} \cdot \textbf{t}\right] \cdot d\textbf{n}$. 
This surface tension, $\gamma_{rot.}$ shows broadly similar trends, see Fig.\,\ref{app:fig:local}\,(b) supporting the importance of including local surface tension changes in the nanoscale region around a rupture.
Whether such local variation of $\gamma$ is consistent across thicknesses, a relatively thicker film with $h_0 \sim 26$ was simulated, and, in accordance to our understanding, a decrement in $\gamma$ was observed, see Fig.\,\ref{app:fig:local}\,(c). However, the decrease in tip surface tension, $\Delta \gamma$ for $h_0 \sim 26$ is less than that for $h_0 \sim 14$. 

\begin{figure}
\centering
\includegraphics[width=0.6\linewidth]{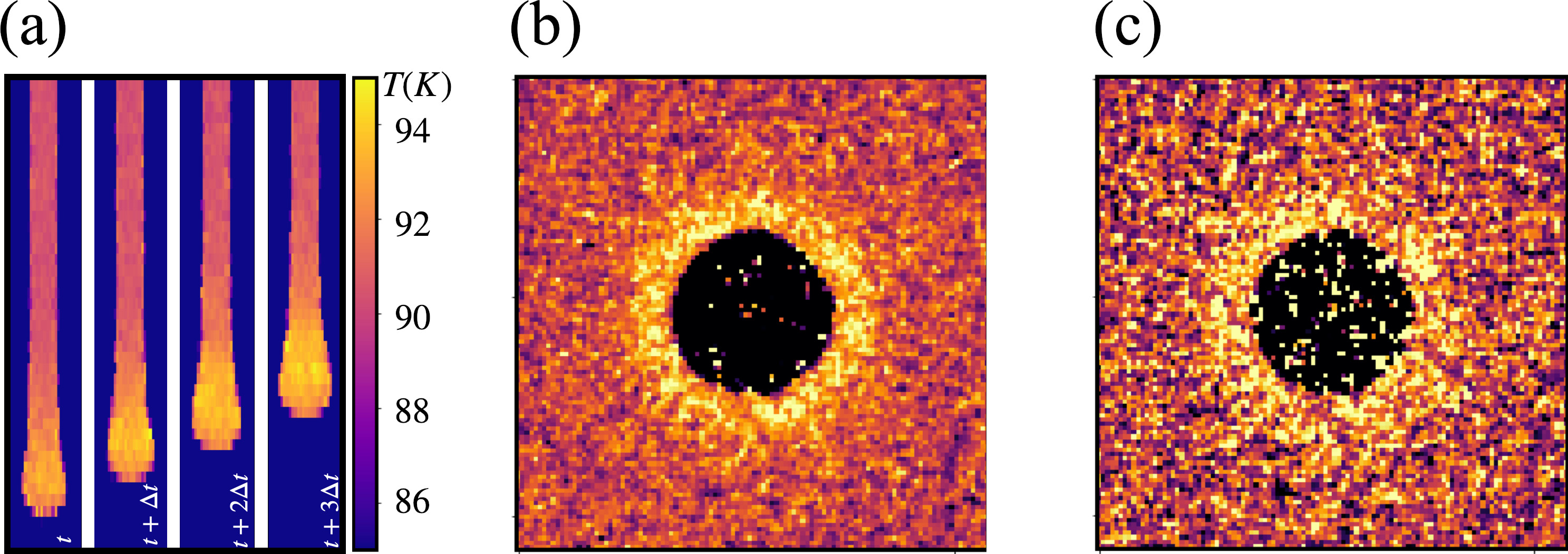}
\caption{(a) Radial-averaged temperature field shows warmer rim/blob region as the film retracts. Top-down view of the film (b) averaged over the thickness, and (c) central-slice shows a region of elevated temperature surrounding the film. For each of the figures, same color-bar applies which is in the units of Kelvin.}
\label{app:fig:temp-qual}
\end{figure}

\begin{figure*}
\centering
\includegraphics[width=0.98\linewidth]{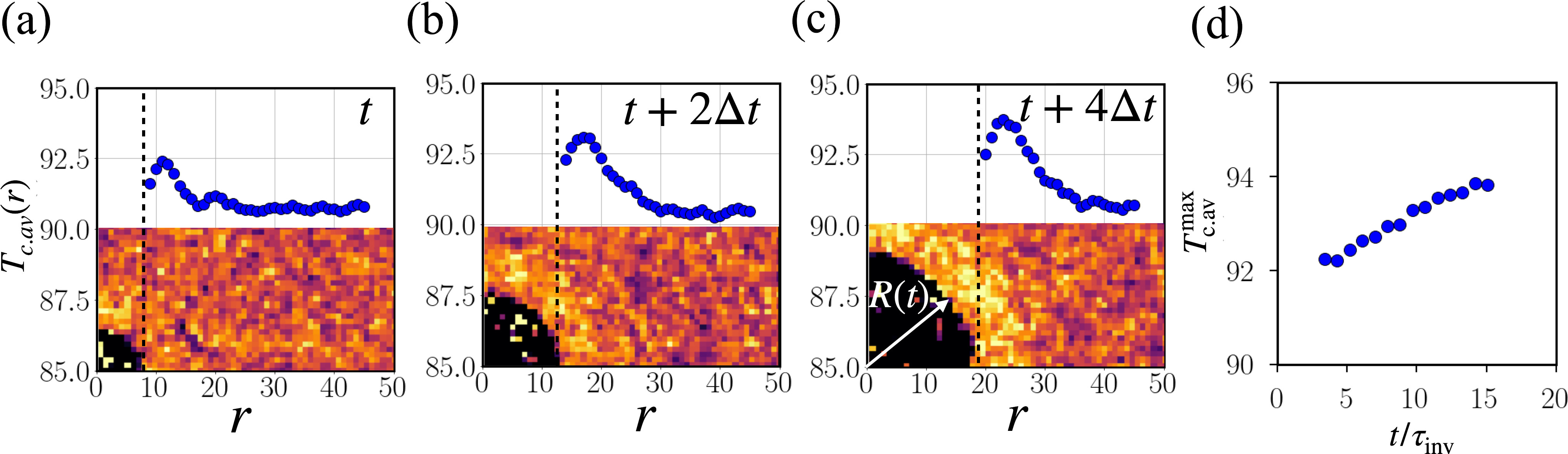}
\caption{(a-c) Averaged temperature along a circular strip of width $dr$ surrounding the hole is plotted against increasing radius. The insets show the top-right quadrant of the corresponding thickness-averaged temperature fields. (d) Maximum temperature (averaged over $\pm10$ time-steps) slowly approaches to convergence. For each of the figures, $Y-$axis is in the units of Kelvin.}
\label{app:fig:temp-quant}
\end{figure*}

{\textbf{Local temperature rise:}}
Fig.\,\ref{app:fig:temp-qual} (a) shows the radial-averaged temperature field for $h_0 \sim 14$, at four instantaneous time as the film retracts. It is observed that the temperature inside the rim is higher than the rest of the film. Similar observations are made from the thickness averaged (panel b) and the central slice (panel c) top-down view of the temperature field. Clearly a warmer rim region surrounds the expanding hole. 
To quantify the temperature increase, an annular region of inner radius, $r_i$, and width $dr$ is drawn which surrounds the hole, and the average temperature of this circular strip is defined as $T_\mathrm{c.av}$. For each instantaneous time, $T_\mathrm{c.av}$ is measured for increasing $r_i$ and plotted against the radial distance from the hole center. Fig.\,\ref{app:fig:temp-quant}\,(a-c) plots $T_\mathrm{c.av}$ against $r$ for three such instantaneous time. Temperature is seen to increase in the part of the film that just surrounds the hole, reaches a peak and then decreases to the film temperature. Notably, and as also observed by \citet{kono2014elevation}, the maximum temperature does not remain constant, rather increases with time (panels a to c). Fig.\,\ref{app:fig:temp-quant}\,(d) plots the temporal evolution of the maximum local temperature, $T_\mathrm{c.av}^\mathrm{max}$.
As seen from the figure and in line with \citet{kono2014elevation}'s observations, $T_\mathrm{c.av}^\mathrm{max}$ tends to convergence, although does not reach within our simulation time window. Note here, the retraction speed in the case considered for this study reaches the constant TC speed before $t/\tau_\mathrm{inv} = 10$. The rise in temperature decreases for thicker films \citep{kono2014elevation}.

\subsection{2.3 Correction for Transition Regime}
\begin{figure}
\centering
    \includegraphics[width=0.65\linewidth]{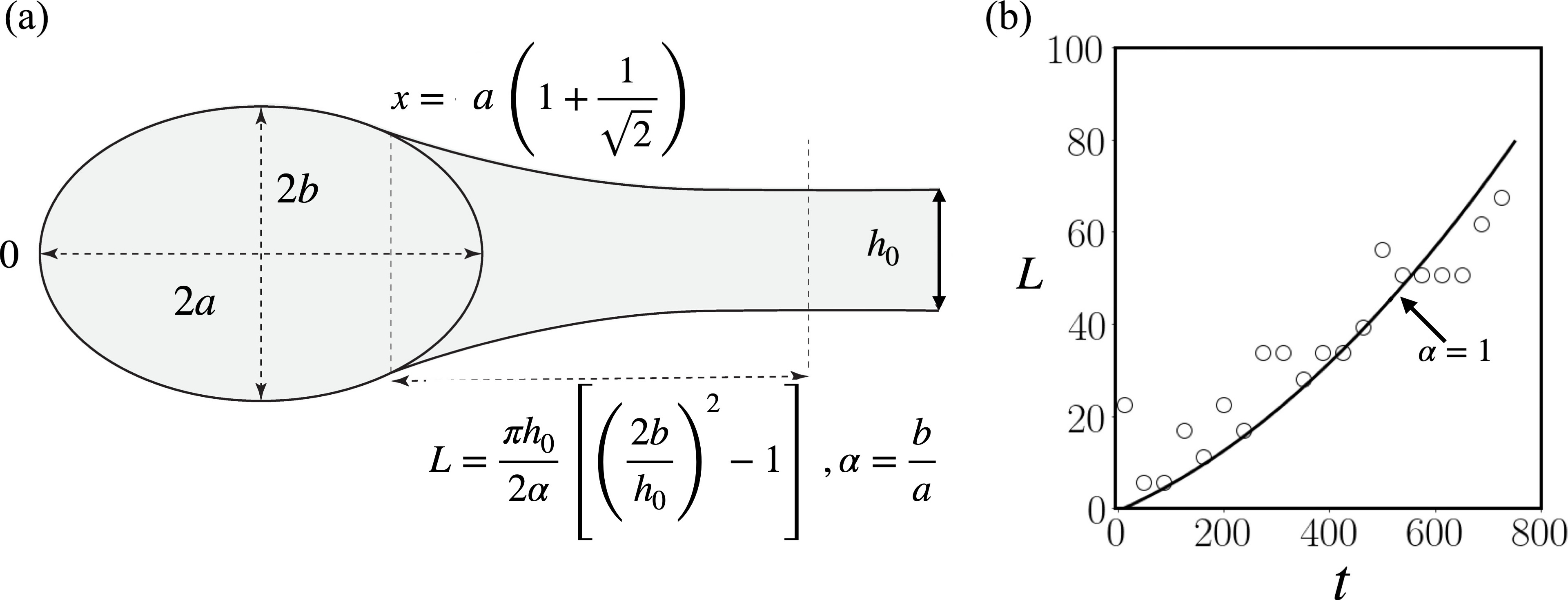}
    \caption{(a) Schematic showing the transition regime between the rim and the unperturbed part of the film, (b) growth of the transition length, $L$ with time. The solid line denotes Eq.\,(\ref{app:eq:tran-len}) for $\alpha = 1$, and symbols are data from simulations.}
    \label{app:fig:transition}
\end{figure}
Results from our simulations show evidence of the existence of a transition region connecting the rim and the unperturbed part of the film. This is schematically shown in Fig.~\ref{app:fig:transition}\,(a), where the rim assumes an elliptic shape with an aspect ratio, $\alpha=b/a$. 
The length of the region of the film that is disturbed by the advancing rim, but not included in the advancing front, can be found from the mass balance. The mass gained by the elliptic blob must be drawn from the undisturbed film. 
Fixing the origin of the co-ordinate system at the tip of the retracting interface, the elliptic rim can be described as: $\frac{(x-a)^2}{a^2}+\frac{y^2}{b^2}=1$.
If the coordinate point, $x=2a$ moves with a constant speed, $2V_w$, one obtains $\frac{\textrm{d}A}{\textrm{d}t}=2\pi V_w b$,
where, $A$ denotes the area of the ellipse. 
This must balance the rate of mass gained by an increase in the transition length $L$, i.e., $h_0 \frac{\textrm{d}L}{\textrm{d}t}=2\pi V_w b$.
Applying the chain rule and integrating with respect to $a$, one obtains $L=\frac{2\pi ab}{h_0}+c$.
Solving for $L=0$ when $2b=h_0$:
\begin{equation}\label{app:eq:tran-len}
\begin{aligned}
        L&=\frac{2\pi ab}{h_0}-\frac{\pi h_0}{2\alpha} \\
        &=\frac{\pi h_0}{2\alpha} \left[\left(\frac{2b}{h_0} \right)^2-1 \right].
\end{aligned}
\end{equation}

Fig.\,\ref{app:fig:transition}\,(b) compares calculated transition length, $L$ from Eq.\,(\ref{app:eq:tran-len}) for $\alpha=1$ with the measured transition length from MD data.
Results from our simulations indicate that the point of transition is where the normal pressure jump is minimised. Applying the Young-Laplace equation, it follows that the absolute value of the curvature must also be minimised. That is, given the form of the ellipse, the meeting point satisfies the condition $y'''=0$. It is straightforward to show that the point of minimisation satisfies $y'=\pm \alpha$,
where $\alpha=b/a$ is the aspect ratio of the ellipse and the negative solution correspond to the upper right quadrant of the ellipse. Since the surface tension on the blob acts parallel to this surface, according to the work of Keller \citep{keller1983breaking}, $h'/2=-\alpha$ and the equation for the velocity becomes:
\begin{equation}
    U= \left(1+\alpha^2\right)^{-1/4} U_c.
\end{equation}
This gives a correction to the Taylor-Culick speed arising from the generalised geometry assumed in the retraction process. 

\bibliography{ref}